\documentclass[aps,prx,10pt,twocolumn,superscriptaddress,showpacs,footinbib,floatfix,subfigure,longbibliography]{revtex4-1}
\usepackage{amsmath,graphics,epstopdf,subfigure,color}

\definecolor{Red}{rgb}{1,0,0}

\def\bra#1{\mathinner{\langle{#1}|}}
\def\ket#1{\mathinner{|{#1}\rangle}}
\def\braket#1{\mathinner{\langle{#1}\rangle}}
\def\ketbra#1#2{{\ket{#1}}{\bra{#2}}}

{\catcode`\|=\active
  \gdef\Braket#1{\begingroup
\mathcode`\|32768\let|\BraVert\left<{#1}\right>\endgroup}}
\def\BraVert{\egroup\,\mid\,\bgroup}
\def\Brak#1#2#3{\bra{#1}#2\ket{#3}}

\begin{document}

	\author{Kavan Modi}
	\email{kavan@quantumlah.org}
	\affiliation{Centre for Quantum Technologies, National University of Singapore, Singapore}

	\author{Hugo Cable}
	\email{cqthvc@nus.edu.sg}
	\affiliation{Centre for Quantum Technologies, National University of Singapore, Singapore}

	\author{Mark Williamson}
	\affiliation{Centre for Quantum Technologies, National University of Singapore, Singapore}

	\author{Vlatko Vedral}
	\affiliation{Centre for Quantum Technologies, National University of Singapore, Singapore}
	\affiliation{Department of Physics, National University of Singapore, Singapore}
	\affiliation{Clarendon Laboratory, University of Oxford, Oxford, UK}
	
\title{Quantum correlations in mixed-state metrology}
\date{\today}

\begin{abstract}
We analyze the effects of quantum correlations, like entanglement and discord, on the efficiency of phase estimation by studying four quantum circuits that can be readily implemented using NMR techniques. These circuits define: a standard strategy of repeated single qubit measurements, a classical strategy where only classical correlations are allowed, and two quantum strategies where nonclassical correlations are allowed. In addition to counting space (number of qubits) and time (number of gates) requirements, we introduce mixedness as a key constraint of the experiment. We compare the efficiency of the four strategies as a function of the mixedness parameter. We find that the quantum strategy gives $\sqrt{N}$ enhancement over the standard strategy for the same amount of mixedness. This result applies even for highly-mixed states that have nonclassical correlations but no entanglement.
\end{abstract}

\maketitle

\section{Introduction}

There is a great deal of work on optimal phase estimation~\cite{natrev, DowlingRev} addressing the practical problems of state generation, particle loss and decoherence. However, this has mainly been done within specific experimental contexts and often with (initially) pure states of the probe only~\cite{natrev}. To understand the origin of the quantum enhancement over the standard quantum limit, many have analyzed the role of the number of bits required and the number of elementary gates needed, as well as the role of entanglement~\cite{PhysRevA.77.052320}. However, in addition to counting the resources, constraints also need to be taken into account. For example, in \emph{nuclear-magnetic-resonance} (NMR)-based quantum information processing, the quantum operations take place at a fixed (room) temperature. This, of course, means that not all physical states can be accessed, only those of a certain (fixed) degree of mixedness. When optimizing phase estimation, this mixedness has to be taken into account. In fact, the degree of mixedness now becomes at least as fundamental as the requirements of the number of qubits and gates.

The other element that plays a crucial role is correlations, namely entanglement when dealing with pure states. However, quantifying correlations as a resource and mixedness as a constraint, leads to a complicated picture. For mixed states entanglement is no longer the sole correlation present; other quantumness quantifiers like quantum discord~\cite{henderson01a, PhysRevLett.88.017901, modietal} may be relevant. A well-studied example of this sort is the \emph{deterministic quantum computation with one qubit} (DQC1)~\cite{KL}. Here, a classically-hard task is performed efficiently quantum mechanically, but no (or only marginal) entanglement is present, while quantum discord can be present even when entanglement is vanishing; this led to the conjecture that discord maybe responsible for the quantum speed-up~\cite{dattashaji}.

In this article, the role of correlations in quantum metrology is studied along the lines of~\cite{dattashaji}. We compare different strategies at a given (fixed) mixedness, within the constraint where pure states are not readily available and classical noise is always present (in contrast to the framework of~\cite{lloydPRL, ToddPRA}). Our study is intended to gain insight into how mixed-state correlations, namely entanglement and discord, contribute to quantum enhancement. We show that mixed-state metrology leads to the same uncertainty in phase estimation as pure states but with an overhead that scales linearly with the classical noise. This turns out to be independent of entanglement, and therefore a quadratic quantum enhancement is available even for states that are highly mixed and fully separable. 

\section{Framework for correlations studies in mixed state metrology}

\begin{figure}
	\begin{center}
\resizebox{5.67 cm}{7.24 cm}{\includegraphics{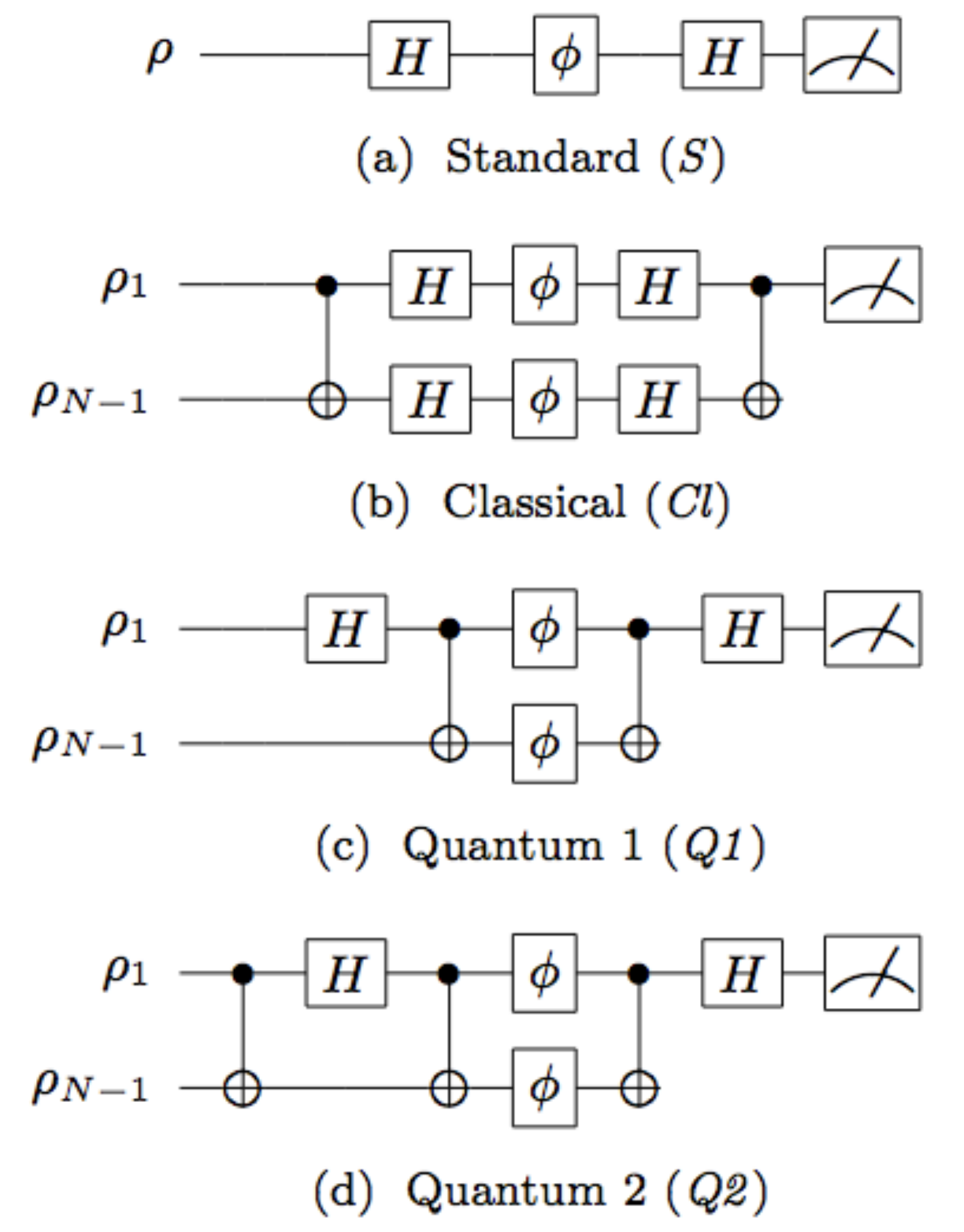}}
\caption{\label{figcircuit} The circuits for the four strategies considered in this paper are shown above. $\rho_1$ is the control qubit and $\rho_{N-1} \equiv \bigotimes_{i=2}^N \rho_i$ are the rest of the $N-1$ qubits. Initially all qubits are in the same state, given in Eq.~\ref{initialstate}.}
	\end{center}
\end{figure}

We work with an $N$-qubit system with each qubit initially being in the mixed state
\begin{align}\label{initialstate}
&\rho=\lambda_0 \ketbra{0}{0} +\lambda_1 \ketbra{1}{1}, \quad \mbox{with}\nonumber\\
&\lambda_0=\frac{1+p}{2} \quad \mbox{and} \quad \lambda_1=\frac{1-p}{2}.
\end{align}
From this we construct correlated states (also called probe states) of various types with unitary gates. Recall that global unitary operations preserve the mixedness of the total state but not the correlations contained within it. We study three strategies having different types of multipartite correlations. The first two are quantum strategies, called \emph{Q1} and \emph{Q2}, which use GHZ-diagonal states. These states have quantum correlations such as entanglement and discord. The third is a classical strategy, labeled \emph{Cl}, which uses only classically-correlated states (defined as having zero discord).  We compare these three strategies to the standard strategy, called \emph{S}, where a single qubit is used $N$ times to estimate the phase. Below we lay out the details of preparing these states.  The circuits for preparing these states are explicitly given in Figs.~\ref{figcircuit}(a)-(d).

\subsection{States preparation}\label{StatesS}

\subsubsection{Standard strategy}

The state for standard strategy is obtained by applying a Hadamard gate, $H$, to each qubit
\begin{gather}\label{standstate}
\varrho_S=(H\rho H)^{\otimes N}.
\end{gather}

\subsubsection{Classical strategy} 

The classical state is created by applying a series of C-Not gates between the first and the $i$th qubit, $C_{1i}$, followed by Hadamard gate on each qubit.
\begin{align}
\varrho_{Cl} =&H_{N} \mathcal{C} \left(\rho_1 \otimes \rho_{N-1} \right) \mathcal{C} H_{N}\nonumber\\
=&\lambda_0\ketbra{+}{+}\otimes(H\rho H)^{\otimes N-1}\nonumber\\
&+\lambda_1\ketbra{-}{-}\otimes(H\sigma_x\rho\sigma_x H)^{\otimes N-1}.
\end{align}
Above $H_{N} \equiv \bigotimes_{i=1}^N H_i$, $\rho_{N-1} \equiv \bigotimes_{i=2}^N \rho_i$ and $\mathcal{C} \equiv \bigotimes_{i=2}^N C_{1i}$, where $C_{1i}$ is a C-Not operation with first qubit as the control and $i$th qubit as the target.

\subsubsection{Quantum strategy 1}

For the first quantum strategy the GHZ diagonal state is prepared by taking the initial uncorrelated $N$ qubit state and applying the Hadamard gate to the first qubit followed by a series of C-Not gates between the first and the $i$th qubit.
\begin{align}
\varrho_{Q1} =&  \mathcal{C} H_1 \left(\rho_1 \otimes \rho_{N-1}\right) H_1 \mathcal{C} \nonumber\\
=&\frac{1}{2}
\begin{pmatrix}
	\rho^{\otimes N-1} & p(\rho\sigma_x)^{\otimes N-1} \\
    p(\sigma_x \rho)^{\otimes N-1} & (\sigma_x \rho\sigma_x)^{\otimes N-1} \\
\end{pmatrix}.\label{ghzstate1}
\end{align}
Above $H_1 \equiv H \otimes \bigotimes_{i=2}^N \openone_i$.

This state was employed in the experiment reported in~\cite{OxfordSci}, but it turns out not to be the optimal state. A more optimal state is described below.

\subsubsection{Quantum strategy 2} 

For the second quantum strategy the GHZ diagonal state is prepared by taking the initial uncorrelated $N$ qubit state and applying C-Not gates between the first and the $i$th qubit followed by the Hadamard gate to the first qubit followed by another series of C-Not gates between the first and the $i$th qubit.
\begin{align}
\varrho_{Q2} =& \mathcal{C} H_{1} \mathcal{C} \left(\rho_1 \otimes \rho_{N-1} \right) \mathcal{C} H_{1} \mathcal{C} \nonumber\\
=&\frac{\lambda_0}{2}
\begin{pmatrix}
	\rho^{\otimes N-1} & (\rho\sigma_x)^{\otimes N-1}\\
	(\sigma_x\rho)^{\otimes N-1} & (\sigma_x\rho\sigma_x)^{\otimes N-1}
\end{pmatrix}\nonumber\\
&+\frac{\lambda_1}{2}
\begin{pmatrix}
	(\sigma_x\rho\sigma_x)^{\otimes N-1} &-(\sigma_x\rho)^{\otimes N-1} \\
	-(\rho\sigma_x)^{\otimes N-1} & \rho^{\otimes N-1}
\end{pmatrix}.\label{ghzstate2}
\end{align}

The \emph{Q2} state is constructed in much of the same way as the \emph{Q1} state, but initialized with C-Not gates to shift the initial population. This strategy was used in the experiment reported in~\cite{simmonsPRA}.

\subsection{Relations to NMR}

Our model corresponds particularly well to a set of recent NMR experiments on magnetic-field sensing~\cite{OxfordSci, simmonsPRA}. The initial state in NMR experiments is `pseudo-pure'---a density matrix which is very close to being completely mixed although its eigenvalues are not quite identical. In NMR, the qubits are the spins of nuclei and unitary operations on these spins are performed by applying electromagnetic pulses of a selected frequency and duration. Qubits can be selectively addressed by choosing spins with specific resonance frequency; these can be local or global (entangling) unitary operations. As more species of spin are added the pulses needed to exclusively address and couple the different species becomes more difficult. However in practice, these operations can be performed with extremely high fidelity (see~\cite{SchaffryPRA} for detailed analysis). 

In the experiments reported in~\cite{OxfordSci, simmonsPRA} only two species were used. This is the so called star topology, where the first qubit ($\rho_1$ in Fig.~\ref{figcircuit}) is used as the control qubit and the rest ($\rho_{N-1}$ in Fig.~\ref{figcircuit}) are subjected to a single transformation at once. For us this translates into using the same one-qubit gate on each of the qubits in $\rho_{N-1}$ and a two-qubit gate between the control qubit and each of the qubits in  $\rho_{N-1}$. The state preparation in~\cite{OxfordSci} slightly differs from the state preparation in~\cite{simmonsPRA}. The difference is precisely the difference in the two quantum strategies considered here: the state in~\cite{OxfordSci} corresponds to the circuit in Fig.~\ref{figcircuit}(c) and the state in~\cite{simmonsPRA} corresponds to the circuit in Fig.~\ref{figcircuit}(d). 

\section{Quantum Fisher information for different strategies}

Now we are in the position to compute the phase uncertainty for each the strategy above. For mixed states, the phase uncertainty is determined by computing the quantum Fisher information~\cite{bcm, LuoLMP, petz} given by
\begin{gather}\label{qfisher}
F(\varrho) =4 \sum_{j> k} \frac{(\eta_j-\eta_k)^2}{\eta_j+\eta_k} |\Brak{\Psi_j}{G}{\Psi_k}|^2,
\end{gather}
where $\{\eta_j\}$ and $\{\ket{\Psi_j}\}$ are the eigenvalues and the corresponding eigenvectors of state $\varrho$, and $G$ is the Hamiltonian of the process that the state is subjected to. The Hamiltonian for the $i$th party is $G_i=\ket{1_i}\bra{1_i}$. For the $N$-party case, each party picks up a phase locally, which means that the global Hamiltonian is $G=\sum_i G_i \otimes \openone_{\bar{i}}$, where the identity matrix acts on the remainder of the Hilbert space. The phase uncertainty is related to quantum Fisher information as
\begin{gather}
\Delta\phi\geq\frac{1}{\sqrt{F(\varrho)}}.
\end{gather}

Quantum Fisher information is a function of the Hamiltonian that generates the interaction between the probe and the object being measured. It also depends on the state of the probe. In our problem, the Hamiltonian is the same for all strategies, only the correlations within the states change. The final measurements at the end are assumed to be optimal generalized measurements as is assumed in the derivation of the quantum Fisher information. For the strategies considered here the measurements turn out to be rather straight forward, see~\cite{SchaffryPRA} for details. The equality in the last equation can be achieved by statistical estimators provided the system is sampled several times. A detailed analysis would identify a statistical estimator to extract the maximum information and saturate the Cram\'er-Rao bound~\cite{BarndorffJPA}. We compute the quantum Fisher information and the phase uncertainties for the three strategies discussed here as follows. At the end of the section we compare these values.

\subsection{Standard strategy}

We begin by computing the quantum Fisher information for $N$ qubits that share no correlations whatsoever.  This is the same as doing the phase estimation experiment with a single qubit and repeating the experiment $N$ times. The initial state of the qubit is taken to be $\varrho_S= (H\rho H)^{\otimes N}.$ The eigenvectors are $\ket{\Psi_{j}}=\ket{r_1,\dots,r_N},$ where $\ket{r_i}\in \{\ket{+}, \ket{-}\}$ is the eigenstate of the $i$th subsystem. We denote an arbitrary degenerate eigenvector, having $\binom{N}{m}$-fold degeneracy, as $\ket{\psi_m}$: Label $m$ counts the number of subsystems in state $\ket{-}$. The corresponding eigenvalue is $\eta_m = \lambda_0^{N-m} \lambda_1^{m}$.

Now, we label the eigenvectors of qubits 2 to $N$ by $\ket{\chi}$ and consider the eigenvectors $\ket{\psi_{m}} = \ket{+,\chi_{m}}$ and $\ket{\psi_{m+1}} = \ket{-,\chi_{m}}$ and the action of the Hamiltonian on them $\braket{\psi_{m}| G |\psi_{m+1}} =\sum_i \braket{+,\chi_{m}| G_i |-,\chi_{m}}$. The only term that survives is $ \braket{+| G_1 |-} \braket{\chi_{m} |\chi_{m}}=\frac{1}{2}.$ Since the states are in the product form, the same result is true for all subsystems and the quantum Fisher information for $N$ qubits is $N$ times the quantum Fisher information of a single qubit
\begin{align}\label{qfisherS}
F(\varrho_S) =&
\sum_{m=0}^{N-1} N \binom{N-1}{m}\nonumber\\
&\times \frac{\left(\lambda_0^{N-m-1} \lambda_1^{m+1} -\lambda_0^{N-m} \lambda_1^{m}\right)^2} {\lambda_0^{N-m-1} \lambda_1^{m+1}+\lambda_0^{N-m} \lambda_1^{m}}\nonumber\\
=& N p^2.
\end{align}
This is the expected result and agrees with the pure state results as $p\rightarrow 1$.

\subsection{Classical strategy}

To create a classical state we start with $\rho^{\otimes N}$, which has eigenvectors $\ket{\Psi_j} =\ket{r_1,\dots,r_N},$ where $\ket{r_i} =\ket{0}, \ket{1}$ is the eigenstate of the $i$th subsystems. Once again we denote an arbitrary degenerate eigenvector, having $\binom{N}{m}$-fold degeneracy, as $\ket{\psi_m}$: Label $m$ counts the number of subsystems in state $\ket{1}$. The corresponding eigenvalue is $\eta_m = \lambda_0^{N-m} \lambda_1^{m}$. 

Next we apply the C-Not gate between the first and the $i$th qubit.  The eigenstates under the C-Not operation change as following: $\ket{\psi_m}= \ket{0,\chi_{m}} \rightarrow \ket{0,\chi_{m}}$ and $\ket{\psi_{m+1}}= \ket{1,\chi_{m}} \rightarrow \ket{\psi_{N-m}}=\ket{1,\chi_{N-m-1}}$. Next a Hadamard gate is applied on each qubit, which simply changes $\ket{0}\rightarrow\ket{+}$ and $\ket{1}\rightarrow\ket{-}$. The action of the Hamiltonian on the eigenstates $\ket{+,\chi_m}$ and $\ket{-,\chi_{m}}$ gives $\braket{+,\chi_m|G|-,\chi_{m}}=\frac{1}{2}$ with the corresponding left and right eigenvalues $\eta_{l}= \lambda_0^{N-m} \lambda_1^{m}$ and $\eta_{r}= \lambda_0^{m} \lambda_1^{N-m}$. $\ket{\chi_m}$ occur with binomial distribution $\binom{N-1}{m}$.

The action of the Hamiltonian on the eigenstates $\ket{\pm,+_i,\chi_{m,\bar{i}}}$ and $\ket{\pm,-_i,\chi_{m,\bar{i}}}$ is $\braket{\pm, +_i, \chi_{m,\bar{i}} |G| \pm, -_i, \chi_{m,\bar{i}}} = \braket{+_i|G_i|-_i} =\frac{1}{2}$ with the $i$th state on the left is $\ket{+}$ and the state on the right is $\ket{-}$. $\ket{\chi_{m,\bar{i}}}$ is the state of parties excluding the first and the $i$th qubits occurring $\binom{N-2}{m}$ times. The index $i$ runs up to $N-1$ yielding the same inner-product.

When the first qubit is in state $\ket{+}$ the corresponding left and right eigenvalues are $\eta_l =\lambda_0^{N-m}\lambda_1^{m}$ and $\eta_{r}= \lambda_0^{N-m-1}\lambda_1^{m+1}$. The difference of these eigenvalues squared divided by their sum is simply $\lambda_0^{N-m-1} \lambda_1^m p^2$. When the first qubit is in state $\ket{-}$ the corresponding left and right eigenvalues are $\eta_l =\lambda_0^{m}\lambda_1^{N-m}$ and $\eta_{r}= \lambda_0^{m+1} \lambda_1^{N-m-1}$.  The difference of these eigenvalues squared divided by their sum is simply $\lambda_0^{m} \lambda_1^{N-m-1} p^2$.

Due to symmetry all other Hamiltonians will have the same result as above. The Fisher information is simply the sum of the three results above

\begin{align}\label{qfisherCl}
F(\varrho_{Cl}) =& \sum_{m=0}^{N-1} \binom{N-1}{m} 
\frac{\left(\lambda_0^{m} \lambda_1^{N-m} -\lambda_0^{N-m} \lambda_1^{m}\right)^2} {\lambda_0^{m} \lambda_1^{N-m}+\lambda_0^{N-m} \lambda_1^{m}}
\nonumber\\
&+ p^2 (N-1) \sum_{m=0}^{N-2} \binom{N-2}{m} \nonumber\\
&\times \left[\lambda_0^{N-m-1}\lambda_1^{m} + \lambda_0^{m} \lambda_1^{N-m-1} \right] \nonumber\\
=& Np^2+1-p^2 \nonumber\\
&-\sum_{m=0}^{N-1} \frac{4}{\lambda_0^{-m}\lambda_1^{m-N} 
+ \lambda_0^{m-N} \lambda_1^{-m}} \binom{N-1}{m}\nonumber\\
\approx& Np^2+1-p^2 - e^{-Np^2},
\end{align}
where the last approximation is obtained numerically.

\subsection{Quantum strategy 1}

The eigenstates of $\varrho_{Q1}$ are of the form $\ket{\psi_{m\pm}} = \frac{1}{\sqrt{2}} \left(\ket{0,\chi_m} \pm \ket{1,\sigma_x^{\otimes N-1}\chi_m}\right)$ with the $+$ eigenvalues being $\lambda_0^{N-m} \lambda_1^{m}$ and the $-$ eigenvalues being $\lambda_0^{N-m-1} \lambda_1^{m+1}$ and $m$ denotes the number of subsystems in state $\ket{1}$. Once again the degeneracies follow the binomial distribution. The action of the Hamiltonian is 
\begin{align}
\braket{\psi_{m-} |G| \psi_{m+}} =& \frac{1}{2}(\bra{0,\chi_m} + \bra{1,\sigma_x^{\otimes N-1}\chi_m}) \nonumber\\
& \times G (\ket{0,\chi_m} -\ket{1,\sigma_x^{\otimes N-1}\chi_m}) \nonumber\\
=&\frac{1}{2}(\bra{0,\chi_m} +\bra{1,\sigma_x^{\otimes N-1}\chi_m}) \nonumber\\
& \times (m\ket{0,\chi_m}-(N-m)\ket{1,\sigma_x^{\otimes N-1}\chi_m}) \nonumber\\
=&\frac{1}{2}(2m-N).
\end{align}
The quantum Fisher information is
\begin{align}\label{qFisherGHZ1}
F(\varrho_{Q1})
=&\sum_{m=0}^{N-1}
\frac{\left(\lambda_0^{N-m-1} \lambda_1^{m+1} - \lambda_0^{N-m} \lambda_1^{m}\right)^2}{\lambda_0^{N-m-1} \lambda_1^{m+1} + \lambda_0^{N-m} \lambda_1^{m}}\nonumber\\
&\times (N-2m)^2
\binom{N-1}{m} \nonumber\\
=& p^2N+2p^3(N-1)
+ (N^2-3N+2)p^4.
\end{align}
Once again, the result above satisfies the known result for pure states. Note that the leading term goes as $p^2N$ and $N^2$ term has a pre-factor of $p^4$.

\subsection{Quantum strategy 2}

The eigenstates of $\varrho_{Q2}$ are of the form $\ket{\psi_{m\pm}} = \frac{1}{\sqrt{2}} \left(\ket{0,\chi_m} \pm \ket{1,\sigma_x^{\otimes N-1}\chi_m}\right)$ with the $+$ eigenvalues being $\lambda_0^{N-m} \lambda_1^{m}$ and the $-$ eigenvalues being $\lambda_0^{m} \lambda_1^{N-m}$ and $m=0,\dots, N$ denotes the number of subsystems in  state $\ket{1}$. Note that the eigenstates here are the same as the previous case but the corresponding eigenvalues are different. Once again the degeneracies follow the binomial distribution. The action of the Hamiltonian is same as the previous case. The quantum Fisher information is
\begin{align}\label{qFisherGHZ2}
F(\varrho_{Q2})
=&\sum_{m=0}^{N-1}
\frac{\left(\lambda_0^{N-m} \lambda_1^{m} - \lambda_0^{m} \lambda_1^{N-m}\right)^2}{\lambda_0^{N-m} \lambda_1^{m} + \lambda_0^{m} \lambda_1^{N-m}}\nonumber\\
&\times(N-2m)^2
\binom{N-1}{m}.
\end{align}
Once again, the result above satisfies the known result for pure states. Numerical results indicate that $F(\varrho_{Q2}) \ge p^2 N^2$. This means that the classical noise is the same as the standard case but we have a quadratic enhancement in the number of qubits.

\subsection{Comparison of quantum Fisher information of different strategies}

\begin{table}[!t]
\begin{center}
\begin{tabular}{cl}\hline\hline
Strategy  & Quantum Fisher information 
\\[6pt]\hline
\emph{S}  & $Np^2$ 
\\[6pt]
\emph{Cl} & $(N-1)p^2+1$ 
\\[0pt]
		   & $\;-\sum_{m=0}^{N-1} \frac{4}{\lambda_0^{-m}\lambda_1^{m-N} + \lambda_0^{m-N} \lambda_1^{-m}} \binom{N-1}{m}$  
\\[6pt]
\emph{Q1} & $Np^2+2p^3(N-1)+p^4(N^2-3N+2)$ 
\\[6pt]
\emph{Q2} & $\sum_{m=0}^{N-1} \frac{(N-2m)^2 \left(\lambda_0^{N-m} \lambda_1^{m} - \lambda_0^{m} \lambda_1^{N-m}\right)^2}{\lambda_0^{N-m} \lambda_1^{m} + \lambda_0^{m} \lambda_1^{N-m}} \binom{N-1}{m}$
\\[0pt]
           & $\geq N^2 p^2$ 
\\[6pt] \hline\hline
\end{tabular}
	\caption{\label{table1} {\sl Quantum Fisher information.} The quantum Fisher information gives the lower bound on the phase uncertainty, $\Delta\phi\geq 1/\sqrt{F(\varrho)}$. Note that classical noise in strategies \emph{S} and \emph{Q2} are the same, $p^2$, independent of $N$. The phase uncertainty for each strategy is plotted as a function of $p$ in Fig.~\ref{fishfig}.}
\end{center}
\end{table}
\begin{figure}[t]
\begin{center}
\resizebox{8 cm}{5 cm}{\includegraphics{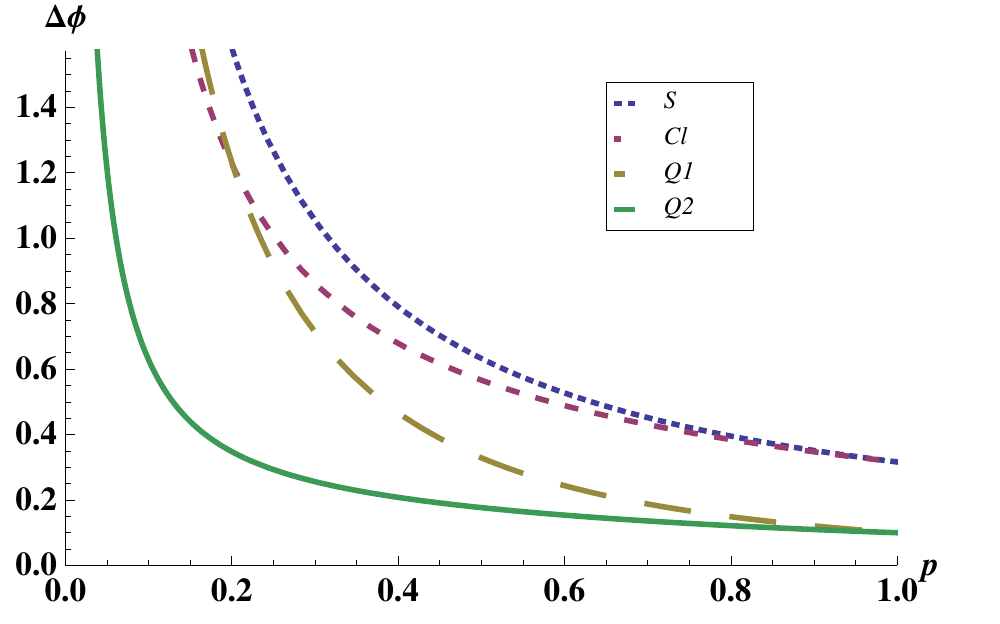}}
 \end{center}
\caption{\label{fishfig}(Color online.) \emph{Phase uncertainties for $N=10$.} The lower bound on the phase uncertainty as given by the quantum Fisher information as functions of $p$ (see Table~\ref{table1}). \emph{Q2} is by far the most optimal strategy for all values of $p$. While, \emph{Cl} strategy is better than \emph{Q1} for small values of $p$.}
\end{figure}

The results of phase uncertainties, presented in the Table~\ref{table1}, are plotted in Fig.~\ref{fishfig} for $N=10$. Remember that our goal is to compare different strategies at a fixed mixedness, i.e. a fixed value of $p$, while changing the number of qubits, i.e. the value of $N$, does not change the overall behavior of these curves. Strategy \emph{Q2} is far better than any of the other strategies, especially \emph{Q1}. In fact, for highly-mixed states the \emph{Cl} strategy is better than \emph{Q1}. The point at which the classical strategy overtakes \emph{Q1} strategy is approximately when $p\approx \frac{1}{\sqrt{N}}$. (This crossing point turns out to be independent of entanglement as the crossing occurs before (for $N=2$) and after (for $N>2$) entanglement vanishes.) Quantum Fisher information for strategy \emph{Q2} is affected by the classical noise in qubits only quadratically, i.e. $F(\varrho_{Q2})\sim (Np)^2$; it could have been exponential in the number of qubits, i.e. $F(\varrho_{Q2})\sim N^2 p^{N}$. Photon losses for optical setups have a devastating effect on quantum enhancement, while the NMR setup seems to be robust under lack of initial coherence.

\section{Optimality and bounds}\label{optimal}

The quantum Fisher information in Eq.~\ref{qfisher} is a function of how the process Hamiltonian can connect two eigenstates of the density matrix and the difference in the corresponding eigenvalues. Maximizing the two will maximize the quantum Fisher information subject to the constraint of the correlation class. Since only unitary operations are allowed for preparation, the spectrum of the density operator remains fixed for all strategies. Therefore the first term of Eq.~\ref{qfisher}, i.e. $(\eta_j - \eta_k)^2 / (\eta_j + \eta_k)$ is fixed. The only change can come from the changes in the eigenvectors. The optimal quantum Fisher information is then given by
\begin{gather}
F_{\max}(\varrho)= 4 \max_{\{U\}} \sum_{j>k} \frac{(\eta_j - \eta_k)^2}{\eta_j + \eta_k} |\braket{\Psi_j | U^\dag G U |\Psi_k}|^2,
\end{gather}
where the unitary transformation $U$ has to be constrained such that it does not change the correlation class of $\varrho$. A rigorous proof of the optimization of the quantum Fisher information for all $N$ and $p$ is a very difficult problem. Below we argue that for any $N$ the states chosen for the strategies \emph{Cl} and \emph{Q2} are optimal for $p$ close to 1 and we conjecture that they remain so for all values of $p$. Certainly they provide strong lower bounds for the quantum Fisher information sufficient to support the conclusions of this article. We should reemphasize these probe states are experimentally realizable and realistic.

\subsection{Optimizing the standard strategy}

For the standard strategy, the quantum Fisher information can be computed for a single party and the $N$-party quantum Fisher information is simply $N$ times the former. The single-party Hamiltonian for the process is $G=\ket{1}\bra{1}$. Therefore the eigenbasis for the density matrix should be $\{\ket{+},\ket{-}\}$ to maximize the transition from one eigenstate to another. This is why the Hadamard gate is applied on all qubits for the preparation.

\subsection{Optimizing the classical strategy}\label{clsopt}

A classical state has a separable (locally orthonormal) eigenbasis~\cite{modietal}, therefore a classical state is simply obtained by rearranging the correspondence between the eigenvectors and the eigenvalues of the $N$-qubit density matrix of the standard strategy (Eq.~\ref{standstate}). Therefore the unitary operations can only permute the computational basis along with local rotations.

The eigenvectors of the classical state are given as $\ket{\Psi_j}=\ket{r_1 \; r_2 \dots r_N}$. The action of the Hamiltonian is 
\begin{align}
G\ket{\Psi_j}=&\sum_{i} G_i \ket{r_1 \; r_2 \dots r_N} \nonumber\\
=& \sum_{i} \braket{1_i| r_i} \ket{r_1 \dots 1_i \dots r_N}. 
\end{align}
For $\braket{\Psi_k|G|\Psi_j}$ to be nonvanishing, $\ket{\Psi_k}$ must only differ from $\ket{\Psi_j}$ only at one site. Then $\braket{\Psi_k|G|\Psi_j}= |\braket{1_i| r_i}|^2$, and the maximum is attained when $\ket{r} \in \{\ket{\pm}\}$. Since the process Hamiltonian is diagonal in $z-$basis, we would like to rotate the eigenvectors to the $x-$basis by applying the Hadamard gate to each qubit, similarly to the standard strategy above.

Now we provide a prescription for maximizing the quantum Fisher information, which is certainly optimal for large $p$, and we conjecture that it remains optimal for all values. The key idea is to insure that the largest and smallest eigenvalues are connected by the process Hamiltonian, i.e. 
\begin{gather}
\max \frac{(\eta_j - \eta_k)^2}{\eta_j + \eta_k}  \quad \forall \quad j,k
\end{gather}
with $\braket{\Psi_j|G|\Psi_k}\neq 0$. The largest eigenvalue is $\lambda_0^N$, belonging to eigenvector $\ket{+\dots +}$. The action of the Hamiltonian is on $\ket{+\dots +}$ is
\begin{gather}
G\ket{+\dots +}=\sum_{i=1}^{N}\frac{1}{\sqrt{2}}\ket{+ \ldots 1_{i} \dots +}.
\end{gather}
This is a superposition of $N$ terms with the $i$th party in state $\ket{1}$. Therefore the only states that have a finite value for $\braket{\psi_m|G|+\ldots+}\neq0$ are: (1) the state with the lowest eigenvalue ($\lambda_1^N$), which becomes
\begin{gather}
\mbox{C-Not}:\ket{--\ldots -}\rightarrow\ket{-+\ldots +};
\end{gather}
(2) states with one excitation, i.e. $\ket{+\ldots -_i \ldots +}$ (there are $N-1$ such states with eigenvalues $\lambda_0^{N-1} \lambda_1$). These latter contributions will be small because the eigenvalues will be different by only one excitation, but occur multiple times. Therefore, $\lambda_0^N - \lambda_1^N$ is the largest possible leading term for the quantum Fisher information. The same argument can be repeated for the eigenvectors with the second largest and the second smallest eigenvalues, and so on, until all eigenvectors are paired in this manner.

\subsection{Optimizing the quantum strategy}

We know for case $p=1$ the optimal pure quantum state for metrology is the GHZ state. For the case $p<1$ we conjecture that a GHZ basis is the optimal basis for quantum Fisher information. Our first attempt along this lines is to use the same circuit as would be used for the pure state case, transforming the eigenstate with the largest eigenvalue, $\lambda_0^N$, into the GHZ state. This in fact strategy \emph{Q1}. The problem with this strategy is that the process Hamiltonian connects this state to a second state that has an eigenvalue that is different by only one excitation, i.e. $\lambda_0^{N-1}\lambda_1$. More precisely, the first term of the quantum Fisher information for strategy \emph{Q1} is
\begin{gather}
\frac{(\lambda_0^N-\lambda_0^{N-1}\lambda_1)^2}{\lambda_0^N+\lambda_0^{N-1}\lambda_1}N^2.
\end{gather}

However, following line of reasoning of the Sec.~\ref{clsopt}, we would like the two eigenvalues for the leading term to be maximum and minimum. Therefore, it would be desirable to permute the eigenvalues of all eigenstates whose leading term is $\ket{1}$. This is precisely what the initial C-Not gates do in \emph{Q2}:
\begin{align}
\mathcal{C} H_1 \mathcal{C} \ket{11\ldots 1} =& \mathcal{C} H_1\ket{10\ldots 0}
=\frac{1}{\sqrt{2}} \mathcal{C} \ket{-0\ldots 0} \nonumber\\
=&\frac{1}{\sqrt{2}}(\ket{00\ldots 0}-\ket{11\ldots 1})
\end{align}
The leading term of quantum Fisher information for strategy \emph{Q2} is then
\begin{gather}
\frac{(\lambda_0^N-\lambda_1^N)^2}{\lambda_0^N+\lambda_1^N}N^2.
\end{gather}

Let us now show that the last term is the largest possible leading term. Suppose that the eigenvector with the largest eigenvalue is connected to some other state not having the smallest eigenvalue and this is the leading term. Explicitly, we have
\begin{gather}
\frac{(\lambda_0^N-\lambda_1^{N-m}\lambda_0^m)^2}{\lambda_0^N+\lambda_1^{N-m}\lambda_0^m}|\braket{\psi_{\mbox{max}}|G|\psi_m}|^2.
\end{gather}
Since the last term in the last equation is independent of $p$, when we take $p=1$ we have
\begin{gather}
|\braket{\psi_{\mbox{max}}|G|\psi_m}|^2\leq N^2.
\end{gather}
$N^2$ is the largest possible value for quantum Fisher information. We also have
\begin{gather}
\frac{(\lambda_0^N-\lambda_1^{N-m}\lambda_0^m)^2}{\lambda_0^N+\lambda_1^{N-m}\lambda_0^m}
\leq\frac{(\lambda_0^N-\lambda_1^{N})^2}{\lambda_0^N+\lambda_1^{N}}
\end{gather}
with equality satisfied if and only if $m=0$ when $p\neq 0,1$. This is true because the numerator becomes smaller and the denominator becomes larger as the value of $m$ increase. Therefore
\begin{gather}
	\frac{(\lambda_0^N-\lambda_1^{N-m}\lambda_0^m)^2}{\lambda_0^N+\lambda_1^{N-m}\lambda_0^m}|\braket{\psi_{\mbox{max}}|G|\psi_m}|^2 \leq 
	\frac{(\lambda_0^N-\lambda_1^N)^2}{\lambda_0^N+\lambda_0^{N-1}\lambda_1^N}N^2
\end{gather}
with equality for $m=0$.

Now that the largest and the smallest eigenvalues are taken care of, we can repeat this process, matching the $m$th smallest eigenvalue with $m$th largest eigenvalue. These arguments strongly suggest that the strategy \emph{Q2} is the optimal quantum strategy for mixed states for all values of $p$ and $N$.

\section{Correlations for different strategies}

In order to relate the results of phase estimation to correlations, we have computed the all of the correlations for the strategies  \emph{Cl}, \emph{Q1}, and \emph{Q2}. The state in the strategy \emph{S} has no correlations and the state in \emph{Cl} strategy does not have any entanglement or discord by definition. Strategies \emph{Q1} and \emph{Q2} have entanglement for some values of $p$, while quantum discord is present for all values of $p$. We begin with computing the entanglement vanishing points for \emph{Q1} and \emph{Q2}.

\subsection{Entanglement in $\varrho_{Q1}$}

It has been shown that a necessary and sufficient condition for a GHZ-diagonal state to be separable is that every possible partial transposition is positive~\cite{Nagata09}. Using this result we can find a relation for a given $N$ that gives the value $p$ for the boundary between separable and entangled. The form of the states we are looking at in the computational basis have already been given in Eq.~\ref{ghzstate1}. Since this matrix is a collection of $2\times 2$ block matrices, the partial transposition that gives the most negative eigenvalues is simply the one that results in a $2\times 2$ matrix with the smallest diagonal elements and the largest off-diagonal elements. Assuming that $p>0$ the $2\times 2$ matrix with the smallest diagonal elements is the one spanning the space in the central part of the matrix with diagonal elements $\lambda_1^{N-1}$. The $2\times 2$ matrix with the largest off-diagonal elements sits in the corners with values $p\lambda_0^{N-1}$. There exists a partial transposition that swaps the smallest off-diagonal elements with the largest off-diagonal elements resulting in the matrix
\begin{gather}
\left(
  \begin{array}{cc}
    \lambda_1^{N-1} & p\; \lambda_0^{N-1} \\
    p\; \lambda_0^{N-1} & \lambda_1^{N-1} \\
  \end{array}
\right).
\end{gather}
The smallest eigenvalue is then $\left(\lambda_1^{N-1}-p\lambda_0^{N-1}\right)$ and it is zero (this is the point the state becomes separable) at $\lambda_1^{N-1} = p \lambda_0^{N-1}.$ One can solve this equation numerically for a given $N$.

\subsection{Entanglement in $\varrho_{Q2}$}

Using the same technique as above and assuming that $p>0$ in Eq.~\ref{ghzstate2} we can show that the $2\times 2$ matrix with the smallest diagonal elements has the elements $\frac{1}{2} \lambda_0^{N_-} \lambda_1^{N_+},$ for $N_{\pm}=N/2$ for even $N$ and $N_{\pm}=(N \pm 1)/2$ for odd $N$. The $2\times 2$ matrix with the largest off-diagonal elements sits in the corners with values $\lambda_0^N-\lambda_1^N$. There exists a partial transposition that swaps the smallest off-diagonal elements with the largest off-diagonal elements resulting in the matrix
\begin{gather}
\left(
  \begin{array}{cc}
    2\lambda_0^{N_-} \lambda_1^{N_+} & \lambda_0^N-\lambda_1^N \\
   \lambda_0^N-\lambda_1^N & 2\lambda_0^{N_-} \lambda_1^{N_+} \\
  \end{array}
\right).
\end{gather}
The smallest eigenvalue is then $\left( 2 \lambda_0^{N_-} \lambda_1^{N_+} -\lambda_0^{N} +\lambda_1^N \right)$, which is zero (this is the point the state becomes separable) at $\lambda_0^{N_-} \lambda_1^{N_+}=\left(\lambda_0^N-\lambda_1^N \right)/2.$ Again, one can solve this equation numerically for a given $N$.

\subsection{Discord in $\varrho_{Q1}$}

\begin{figure}[t]
\begin{center}
\resizebox{7 cm}{4.82 cm}{\includegraphics{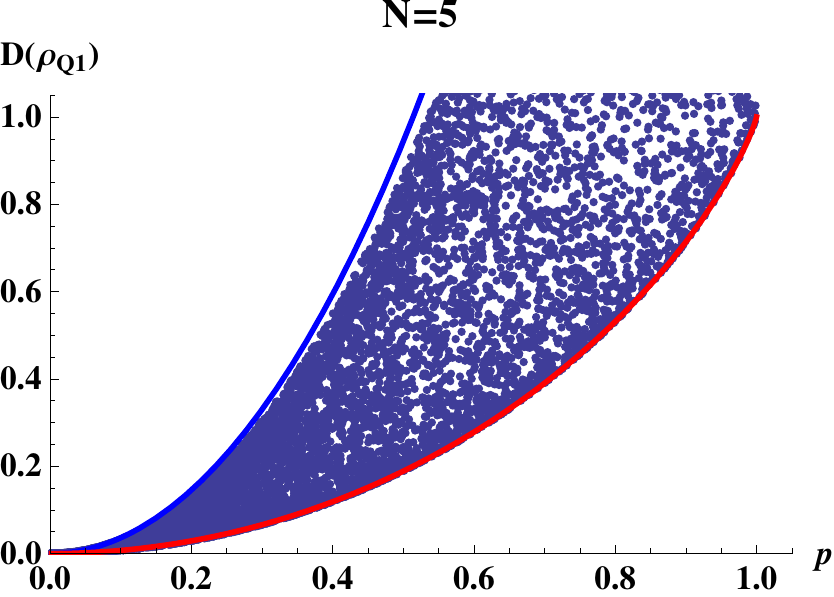}}
\quad\quad\quad\quad
\resizebox{7 cm}{4.82 cm}{\includegraphics{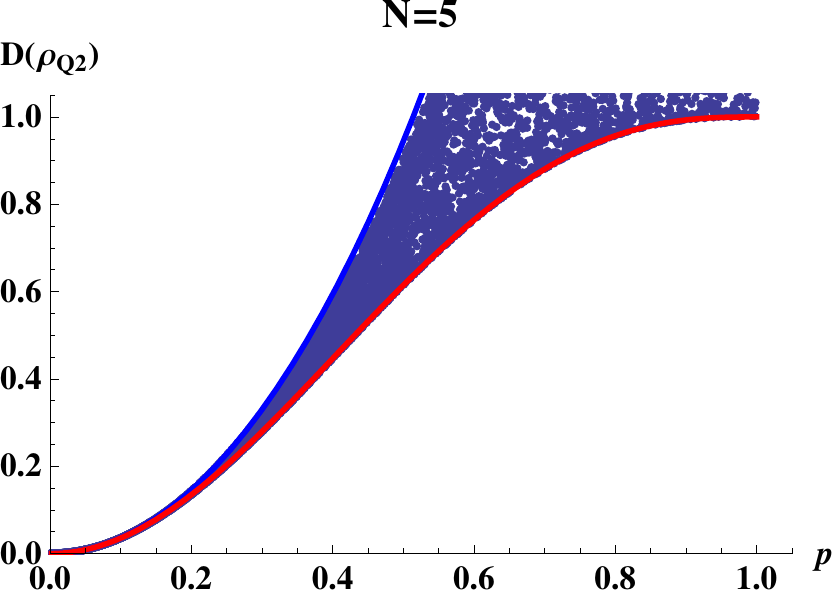}}
   \end{center}
\caption{\label{simdiscord}(Color online.)\emph{Numerical simulation of discord for Q1(Q2).} Simulation of quantum discord ($D$) for $N=5$ with 10,000 random measurements on $\rho_{Q1(Q2)}$ to obtain $\chi_{Q1(Q2)}$. The top (blue) line is the maximum possible values for $D$, i.e. $5-S(\varrho_{Q1(Q2)})$. The bottom (red) line is the conjectured $D=S(\chi_{Q1(Q2)})-S(\varrho_{Q1(Q2)})$. All random measurements points fall between the two lines suggesting that the conjectured formulae to be correct given Table~\ref{tabcor}.}
\end{figure}

Quantum discord, denoted by $D$, is defined as the distance (using relative entropy) between a quantum state and it's closest classical state: $D(\varrho) = \min_{\{\ket{\bf k}\}} S(\varrho||\chi_\varrho)$. The closest classical state, $\chi_\varrho$, is found by dephasing $\varrho$ in a locally-orthonormal product basis $\{\ket{\bf k}\}$: $\chi_\varrho=\sum_{\bf k}\ket{\bf k}\bra{\bf k} \varrho \ket{\bf k}\bra{\bf k}$, (see~\cite{modietal} for details).  We note that discord serves as the upper bound on entanglement as a function of $p$, $E(p)\leq D(p)$~\cite{modietal}. 

Computing quantum discord is an extremely hard problem; there exists no closed form solution even for arbitrary two-qubit states: The main difficulty lies in determining the minimizing basis $\{\ket{\bf k}\}$. In this problem we are dealing with a multi-qubit state. Using relative entropy of discord avoids making arbitrary bipartitions as would be required for computing bipartite discord.

However, following the recipe of~\cite{modietal}, the closest classical state to $\varrho_{Q1}$ is conjectured to be given by dephasing $\varrho_{Q1}$ in the standard basis:
\begin{gather}
\chi_{Q1}=\frac{1}{2}
\begin{pmatrix}
	\rho^{\otimes N-1} & 0 \\
    0 & (\sigma_x \rho\sigma_x)^{\otimes N-1} \\
\end{pmatrix}.
\end{gather}
To calculate $D$ we just need to take the difference in the entropies of $\varrho_{Q1}$ and $\chi_{Q1}$.
\begin{gather}
D_{Q1}=S(\chi_{Q1})-S(\varrho_{Q1})=1-S(\rho),
\end{gather}
where $S$ is the von Neumann entropy: $S(\rho)=-\mbox{tr}[\rho \log(\rho) ]$ and $\rho$ is the state given in Eq.~\ref{initialstate}. 

Since the last equation is a conjecture, we have numerically simulated the closest classical state for up to five qubits (see Fig.~\ref{simdiscord}). The result above holds up (i.e. discord is independent of $N$), but we do not yet have an analytic proof. One can consider this result to be at least an upper bound on discord. We should note that the lower bound discord is strictly greater than 0, as it is easy to verify it is a quantum correlated state~\cite{arXiv:1005.4348}. Finally, we have plotted discord given in the last equation as a function of $p$ in Fig.~\ref{corrfig-a}.

\subsection{Discord in $\varrho_{Q2}$}

\begin{figure}[t!]
\begin{center}
\subfigure[$\;$ Quantum Discord]
{\label{corrfig-a}
\resizebox{8 cm}{5 cm}{\includegraphics{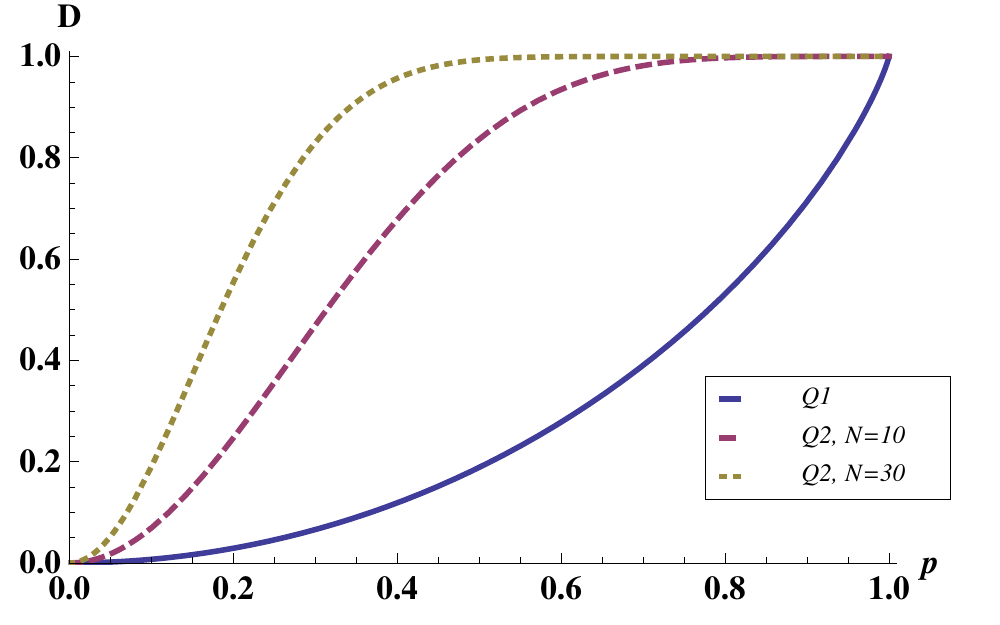}}}
    \subfigure[$\;$ Classical Correlations]
{\label{corrfig-b}
\resizebox{8 cm}{5 cm}{\includegraphics{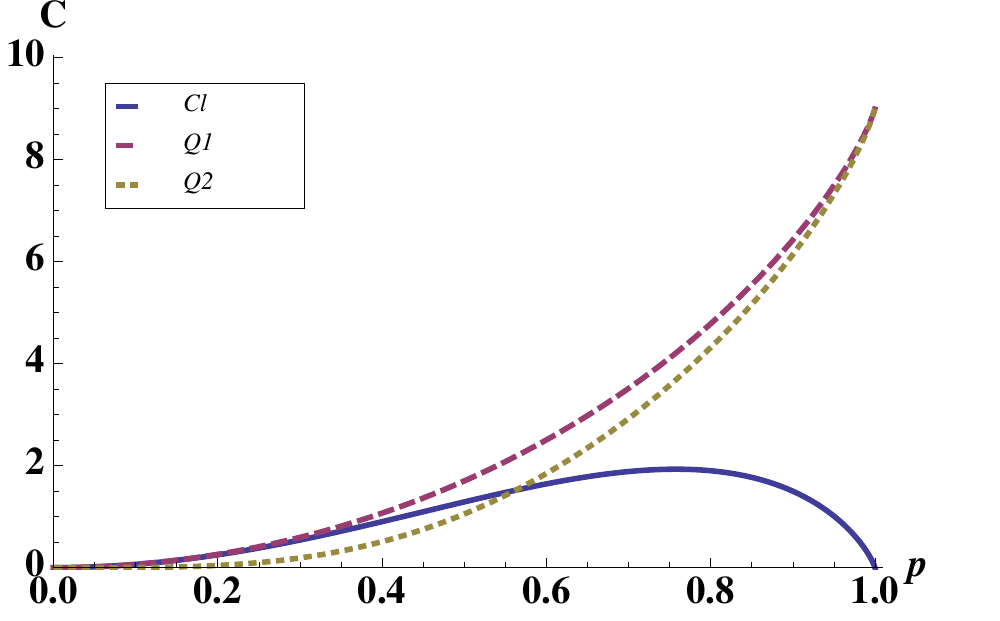}}}
 \end{center}
\caption{(Color online.) \emph{Conjectured discord and classical correlations as functions of $p$.} (a) Discord is always present for the two quantum strategies. Discord in \emph{Q1} is independent of $N$. Entanglement is always (equal at $p=1$ or) smaller than discord and vanishes (for $N=10$) around $p=0.118$ for \emph{Q1} and around $p=0.088$ for \emph{Q2}. (b) Classical correlations for $N=10$ are plotted as function of the mixedness.}
\end{figure}

Since both $\varrho_{Q1}$ and $\varrho_{Q2}$ are GHZ-diagonal states, the form of their closest classical states are also the same. Which means we can simply dephase $\varrho_{Q2}$ in the computation basis to get:
\begin{align}
\chi_{Q2}=&	\frac{\lambda_0}{2}
	\begin{pmatrix}
		\rho^{\otimes N-1} & 0 \\ 0 & (\sigma_x\rho\sigma_x)^{\otimes N-1}
	\end{pmatrix}\nonumber\\
	&+\frac{\lambda_1}{2}
	\begin{pmatrix}
		(\sigma_x\rho\sigma_x)^{\otimes N-1} & 0 \\ 0 & \rho^{\otimes N-1}
	\end{pmatrix}.
\end{align}
To calculate $D$ we just need to take the difference in the entropies of $\varrho_{GHZ}$ and $\chi$.
\begin{align}
D_{Q2}=&S(\chi_{Q2})-S(\varrho_{Q2})\nonumber\\
=&2\sum_m h\left(\frac{\lambda_0^{N-m}\lambda_1^{m}+\lambda_0^{m}\lambda_1^{N-m}}{2}\right)\nonumber\\
&\times{\left(\begin{matrix} N-1\\ m \end{matrix}\right)}-N S(\rho),
\end{align}
where $h(x)=-x \log(x)$.  Once again this formula is conjectured, but numerical evidence shown in Fig.~\ref{simdiscord} supports this result. Finally, we have plotted discord given in the last equation as a function of $p$ in Fig.~\ref{corrfig-a}. 

\begin{table}[t]
\begin{tabular}{l}
\hline\hline
{Entanglement vanishing points}\cr\hline
\begin{tabular}{l}
$E_{Q1}=0 \quad \Leftrightarrow \quad p\leq(\lambda_1/\lambda_0)^{N-1}$\cr
$E_{Q2}=0 \quad \Leftrightarrow \quad \lambda_0^{N_-} \lambda_1^{N_+} \geq\frac{1}{2}\left(\lambda_0^N-\lambda_1^N\right)$\cr
\end{tabular}\cr
{ Quantum discord}\cr\hline
\begin{tabular}{l}
$D_{Q1} = 1-S(\rho)$\cr
$D_{Q2} = 2\sum_{m=0}^{N-1} \binom{N}{m-1}
h\left(\frac{\lambda_0^{N-m}\lambda_1^{m}+\lambda_0^{m}\lambda_1^{N-m}}{2}\right)-N S(\rho)$\cr
\end{tabular}\cr
{ Classical correlations}\cr\hline
\begin{tabular}{l}
$C_{Cl} = (N-1)\left[h\left(\lambda_0^2+\lambda_1^2\right) +h\left( 2\lambda_0 \lambda_1 \right) -S(\rho)\right]$\cr
$C_{Q1} = (N-1)[1-S(\rho)]$\cr
$C_{Q2} = N [1-S(\rho)]-D_{Q2}$\cr
\end{tabular}
\end{tabular}
\caption{\label{tabcor} Above $N_\pm=(N \pm 1)/2$ if $N$ is odd and $N_\pm=N/2$ when $N$ is even. $S(\rho)=-\mbox{tr} [\rho \log (\rho)]$ is the von Neumann entropy and $h(x)=-x \log(x)$. The expressions for discord (and therefore classical correlations) above are conjectured and supported by numerical studies (see Fig.~\ref{simdiscord}). The details of the calculations are below. We have plotted these correlations as functions of $p$ in Figs.~\ref{corrfig-a} and~\ref{corrfig-b}.}
\end{table}

\subsection{Review of correlations in probe states.}

Finally, we compute classical correlations following again the procedure in~\cite{modietal} and shown in Table~\ref{tabcor}. We have plotted the classical correlations given in the last equations as a function of $p$ in Fig.~\ref{corrfig-b}. In Table~\ref{tabcor} we list the formulae for all correlations computed in this section.

\section{Analysis}

Now we are in the position to relate correlations with the enhancement of the quantum Fisher information. We start by noticing that quantum Fisher information is affected by classical noise in qubits only quadratically, i.e. $F(\varrho_{Q2})\sim (Np)^2$, while entanglement for \emph{Q2} vanishes when $p \approx 1/N$. Classical correlations for the three strategies scale linearly with the number of qubits $N$ (see Table.~\ref{tabcor}). In fact, \emph{Q1} has more classical correlations than \emph{Cl} and \emph{Q2} for all values of $p$. This supports the expected result that classical correlations, although, present in bulk do not contribute to quantum enhancement. The total correlations, defined as the sum of quantum discord and classical correlations: $T=D+C$, are the same for both \emph{Q1} and \emph{Q2}. This further allows us to distinguishes the role of quantum correlations in the two cases. 

Which brings us to our main observations. The enhancement of phase uncertainty (hence quantum Fisher information) due to the optimal quantum strategy over the standard strategy is
\begin{gather}
\mbox{Quantum Advantage}=\sqrt{\frac{F_{Q2}}{F_{S}}}\approx\sqrt{N} \quad \forall\;p.
\end{gather}
Since the classical noise is roughly $p^2$ for both strategies \emph{S} and \emph{Q2} (see Table~\ref{table1}), the quantum advantage is $\sqrt{N}$. This is true for highly-mixed states that have no entanglement, i.e. $p$ close to zero. Surprisingly, not a great deal of quantum coherence is needed to attain quantum advantage in quantum metrology.

For the experiments reported in~\cite{OxfordSci,simmonsPRA} $p\approx 10^{-5}$ and $N\approx 10$. Therefore both states $\varrho_{Q1}$ or $\varrho_{Q2}$ are unentangled. Both experiments reported quantum enhancement, which is in accordance with our findings. Quantum discord, on the other hand, does not vanish until $p\rightarrow 0$. And for \emph{Q2}, quantum discord depends on the number of qubits, unlike for \emph{Q1} (see Table.~\ref{tabcor}), assuming our conjectured expressions for discord are fully valid. Quantum discord for \emph{Q2} grows for small values of $p$ as $N$ increases. This provides evidence that quantum discord may have some responsibility for the enhancement in quantum metrology. We should note that for entanglement to appear when $p \approx 10^{-5}$, the number of qubits has to be roughly $N \approx 10^5$.

In conclusion we have analyzed the role of quantum and classical correlations in mixed-state phase estimation. We found evidence that classical correlations do not play a large role in quantum enhancement, as expected. However, we also showed that quadratic quantum enhancement does not vanish as entanglement vanishes. For such states quantum discord is present and is a growing function of the number of qubits. This adds to the evidence that quantum discord may be responsible for some quantum enhancements.

\acknowledgements We acknowledge the financial support by the National Research Foundation and the Ministry of Education of Singapore. We thank E. Gauger, V. Giovannetti, J. Jones, B. Lovett, B. Munro, K. Nemoto, M. Schaffry, S. Simmons, T. Tilma, for helpful conversations.

\bibliography{metrology.bib}

\begin{thebibliography}{19}%
\makeatletter
\providecommand \@ifxundefined [1]{%
 \@ifx{#1\undefined}
}%
\providecommand \@ifnum [1]{%
 \ifnum #1\expandafter \@firstoftwo
 \else \expandafter \@secondoftwo
 \fi
}%
\providecommand \@ifx [1]{%
 \ifx #1\expandafter \@firstoftwo
 \else \expandafter \@secondoftwo
 \fi
}%
\providecommand \natexlab [1]{#1}%
\providecommand \enquote  [1]{``#1''}%
\providecommand \bibnamefont  [1]{#1}%
\providecommand \bibfnamefont [1]{#1}%
\providecommand \citenamefont [1]{#1}%
\providecommand \href@noop [0]{\@secondoftwo}%
\providecommand \href [0]{\begingroup \@sanitize@url \@href}%
\providecommand \@href[1]{\@@startlink{#1}\@@href}%
\providecommand \@@href[1]{\endgroup#1\@@endlink}%
\providecommand \@sanitize@url [0]{\catcode `\\12\catcode `\$12\catcode
  `\&12\catcode `\#12\catcode `\^12\catcode `\_12\catcode `\%12\relax}%
\providecommand \@@startlink[1]{}%
\providecommand \@@endlink[0]{}%
\providecommand \url  [0]{\begingroup\@sanitize@url \@url }%
\providecommand \@url [1]{\endgroup\@href {#1}{\urlprefix }}%
\providecommand \urlprefix  [0]{URL }%
\providecommand \Eprint [0]{\href }%
\providecommand \doibase [0]{http://dx.doi.org/}%
\providecommand \selectlanguage [0]{\@gobble}%
\providecommand \bibinfo  [0]{\@secondoftwo}%
\providecommand \bibfield  [0]{\@secondoftwo}%
\providecommand \translation [1]{[#1]}%
\providecommand \BibitemOpen [0]{}%
\providecommand \bibitemStop [0]{}%
\providecommand \bibitemNoStop [0]{.\EOS\space}%
\providecommand \EOS [0]{\spacefactor3000\relax}%
\providecommand \BibitemShut  [1]{\csname bibitem#1\endcsname}%
\let\auto@bib@innerbib\@empty
\bibitem [{\citenamefont {Giovannetti}\ \emph {et~al.}(2011)\citenamefont
  {Giovannetti}, \citenamefont {Lloyd},\ and\ \citenamefont
  {Maccone}}]{natrev}%
  \BibitemOpen
  \bibfield  {author} {\bibinfo {author} {\bibfnamefont {V.}~\bibnamefont
  {Giovannetti}}, \bibinfo {author} {\bibfnamefont {S.}~\bibnamefont {Lloyd}},
  \ and\ \bibinfo {author} {\bibfnamefont {L.}~\bibnamefont {Maccone}},\
  }\bibfield  {title} {\enquote {\bibinfo {title} {Advances in quantum
  metrology},}\ }\href@noop {} {\bibfield  {journal} {\bibinfo  {journal}
  {Nature Photonics}\ }\textbf {\bibinfo {volume} {5}},\ \bibinfo {pages} {222}
  (\bibinfo {year} {2011})}\BibitemShut {NoStop}%
\bibitem [{\citenamefont {Dowling}(2008)}]{DowlingRev}%
  \BibitemOpen
  \bibfield  {author} {\bibinfo {author} {\bibfnamefont {J.}~\bibnamefont
  {Dowling}},\ }\bibfield  {title} {\enquote {\bibinfo {title} {Quantum optical
  metrology -- the lowdown on high-{N00N} states},}\ }\href@noop {} {\bibfield
  {journal} {\bibinfo  {journal} {Contemp. Phys.}\ }\textbf {\bibinfo {volume}
  {49}},\ \bibinfo {pages} {125--143} (\bibinfo {year} {2008})}\BibitemShut
  {NoStop}%
\bibitem [{\citenamefont {Boixo}\ and\ \citenamefont
  {Somma}(2008)}]{PhysRevA.77.052320}%
  \BibitemOpen
  \bibfield  {author} {\bibinfo {author} {\bibfnamefont {Sergio}\ \bibnamefont
  {Boixo}}\ and\ \bibinfo {author} {\bibfnamefont {Rolando~D.}\ \bibnamefont
  {Somma}},\ }\bibfield  {title} {\enquote {\bibinfo {title} {Parameter
  estimation with mixed-state quantum computation},}\ }\href {\doibase
  10.1103/PhysRevA.77.052320} {\bibfield  {journal} {\bibinfo  {journal} {Phys.
  Rev. A}\ }\textbf {\bibinfo {volume} {77}},\ \bibinfo {pages} {052320}
  (\bibinfo {year} {2008})}\BibitemShut {NoStop}%
\bibitem [{\citenamefont {Henderson}\ and\ \citenamefont
  {Vedral}(2001)}]{henderson01a}%
  \BibitemOpen
  \bibfield  {author} {\bibinfo {author} {\bibfnamefont {L.}~\bibnamefont
  {Henderson}}\ and\ \bibinfo {author} {\bibfnamefont {V.}~\bibnamefont
  {Vedral}},\ }\bibfield  {title} {\enquote {\bibinfo {title} {Classical,
  quantum and total correlations},}\ }\href@noop {} {\bibfield  {journal}
  {\bibinfo  {journal} {J. Phys. A}\ }\textbf {\bibinfo {volume} {34}},\
  \bibinfo {pages} {6899} (\bibinfo {year} {2001})}\BibitemShut {NoStop}%
\bibitem [{\citenamefont {Ollivier}\ and\ \citenamefont
  {Zurek}(2001)}]{PhysRevLett.88.017901}%
  \BibitemOpen
  \bibfield  {author} {\bibinfo {author} {\bibfnamefont {H.}~\bibnamefont
  {Ollivier}}\ and\ \bibinfo {author} {\bibfnamefont {W.~H.}\ \bibnamefont
  {Zurek}},\ }\bibfield  {title} {\enquote {\bibinfo {title} {Quantum discord:
  {A} measure of the quantumness of correlations},}\ }\href {\doibase
  10.1103/PhysRevLett.88.017901} {\bibfield  {journal} {\bibinfo  {journal}
  {Phys. Rev. Lett.}\ }\textbf {\bibinfo {volume} {88}},\ \bibinfo {pages}
  {017901} (\bibinfo {year} {2001})}\BibitemShut {NoStop}%
\bibitem [{\citenamefont {Modi}\ \emph {et~al.}(2010)\citenamefont {Modi},
  \citenamefont {Paterek}, \citenamefont {Son}, \citenamefont {Vedral},\ and\
  \citenamefont {Williamson}}]{modietal}%
  \BibitemOpen
  \bibfield  {author} {\bibinfo {author} {\bibfnamefont {Kavan}\ \bibnamefont
  {Modi}}, \bibinfo {author} {\bibfnamefont {Tomasz}\ \bibnamefont {Paterek}},
  \bibinfo {author} {\bibfnamefont {Wonmin}\ \bibnamefont {Son}}, \bibinfo
  {author} {\bibfnamefont {Vlatko}\ \bibnamefont {Vedral}}, \ and\ \bibinfo
  {author} {\bibfnamefont {Mark}\ \bibnamefont {Williamson}},\ }\bibfield
  {title} {\enquote {\bibinfo {title} {Unified view of quantum and classical
  correlations},}\ }\href@noop {} {\bibfield  {journal} {\bibinfo  {journal}
  {Phys. Rev. Lett.}\ }\textbf {\bibinfo {volume} {104}},\ \bibinfo {pages}
  {080501} (\bibinfo {year} {2010})}\BibitemShut {NoStop}%
\bibitem [{\citenamefont {Knill}\ and\ \citenamefont {Laflamme}(1998)}]{KL}%
  \BibitemOpen
  \bibfield  {author} {\bibinfo {author} {\bibfnamefont {E.}~\bibnamefont
  {Knill}}\ and\ \bibinfo {author} {\bibfnamefont {R.}~\bibnamefont
  {Laflamme}},\ }\bibfield  {title} {\enquote {\bibinfo {title} {Power of one
  bit of quantum information},}\ }\href@noop {} {\bibfield  {journal} {\bibinfo
   {journal} {Phys. Rev. Lett.}\ }\textbf {\bibinfo {volume} {81}},\ \bibinfo
  {pages} {5672} (\bibinfo {year} {1998})}\BibitemShut {NoStop}%
\bibitem [{\citenamefont {Datta}\ \emph {et~al.}(2008)\citenamefont {Datta},
  \citenamefont {Shaji},\ and\ \citenamefont {Caves}}]{dattashaji}%
  \BibitemOpen
  \bibfield  {author} {\bibinfo {author} {\bibfnamefont {A.}~\bibnamefont
  {Datta}}, \bibinfo {author} {\bibfnamefont {A.}~\bibnamefont {Shaji}}, \ and\
  \bibinfo {author} {\bibfnamefont {C.}~\bibnamefont {Caves}},\ }\bibfield
  {title} {\enquote {\bibinfo {title} {Quantum discord and the power of one
  qubit},}\ }\href@noop {} {\bibfield  {journal} {\bibinfo  {journal} {Phys.
  Rev. Lett.}\ }\textbf {\bibinfo {volume} {100}},\ \bibinfo {pages} {050502}
  (\bibinfo {year} {2008})}\BibitemShut {NoStop}%
\bibitem [{\citenamefont {Giovannetti}\ \emph {et~al.}(2006)\citenamefont
  {Giovannetti}, \citenamefont {Lloyd},\ and\ \citenamefont
  {Maccone}}]{lloydPRL}%
  \BibitemOpen
  \bibfield  {author} {\bibinfo {author} {\bibfnamefont {V.}~\bibnamefont
  {Giovannetti}}, \bibinfo {author} {\bibfnamefont {S.}~\bibnamefont {Lloyd}},
  \ and\ \bibinfo {author} {\bibfnamefont {L.}~\bibnamefont {Maccone}},\
  }\bibfield  {title} {\enquote {\bibinfo {title} {Quantum metrology},}\
  }\href@noop {} {\bibfield  {journal} {\bibinfo  {journal} {Phys. Rev. Lett.}\
  }\textbf {\bibinfo {volume} {96}},\ \bibinfo {pages} {010401} (\bibinfo
  {year} {2006})}\BibitemShut {NoStop}%
\bibitem [{\citenamefont {Tilma}\ \emph {et~al.}(2010)\citenamefont {Tilma},
  \citenamefont {Hamaji}, \citenamefont {Munro},\ and\ \citenamefont
  {Nemoto}}]{ToddPRA}%
  \BibitemOpen
  \bibfield  {author} {\bibinfo {author} {\bibfnamefont {Todd}\ \bibnamefont
  {Tilma}}, \bibinfo {author} {\bibfnamefont {Shinichiro}\ \bibnamefont
  {Hamaji}}, \bibinfo {author} {\bibfnamefont {W.~J.}\ \bibnamefont {Munro}}, \
  and\ \bibinfo {author} {\bibfnamefont {Kae}\ \bibnamefont {Nemoto}},\
  }\bibfield  {title} {\enquote {\bibinfo {title} {Entanglement is not a
  critical resource for quantum metrology},}\ }\href {\doibase
  10.1103/PhysRevA.81.022108} {\bibfield  {journal} {\bibinfo  {journal} {Phys.
  Rev. A}\ }\textbf {\bibinfo {volume} {81}},\ \bibinfo {pages} {022108}
  (\bibinfo {year} {2010})}\BibitemShut {NoStop}%
\bibitem [{\citenamefont {Jones}\ \emph {et~al.}(2009)\citenamefont {Jones},
  \citenamefont {Karlen}, \citenamefont {Fitzsimons}, \citenamefont {Ardavan},
  \citenamefont {Benjamin}, \citenamefont {Briggs},\ and\ \citenamefont
  {Morton}}]{OxfordSci}%
  \BibitemOpen
  \bibfield  {author} {\bibinfo {author} {\bibfnamefont {J.~A.}\ \bibnamefont
  {Jones}}, \bibinfo {author} {\bibfnamefont {S.~D.}\ \bibnamefont {Karlen}},
  \bibinfo {author} {\bibfnamefont {J.}~\bibnamefont {Fitzsimons}}, \bibinfo
  {author} {\bibfnamefont {A.}~\bibnamefont {Ardavan}}, \bibinfo {author}
  {\bibfnamefont {S.~C.}\ \bibnamefont {Benjamin}}, \bibinfo {author}
  {\bibfnamefont {G.~A.~D.}\ \bibnamefont {Briggs}}, \ and\ \bibinfo {author}
  {\bibfnamefont {J.~J.~L.}\ \bibnamefont {Morton}},\ }\bibfield  {title}
  {\enquote {\bibinfo {title} {{Magnetic field sensing beyond the standard
  quantum limit using 10-spin NOON states}},}\ }\href@noop {} {\bibfield
  {journal} {\bibinfo  {journal} {Science}\ }\textbf {\bibinfo {volume}
  {324}},\ \bibinfo {pages} {1166--1168} (\bibinfo {year} {2009})}\BibitemShut
  {NoStop}%
\bibitem [{\citenamefont {Simmons}\ \emph {et~al.}(2010)\citenamefont
  {Simmons}, \citenamefont {Jones}, \citenamefont {Karlen}, \citenamefont
  {Ardavan},\ and\ \citenamefont {Morton}}]{simmonsPRA}%
  \BibitemOpen
  \bibfield  {author} {\bibinfo {author} {\bibfnamefont {Stephanie}\
  \bibnamefont {Simmons}}, \bibinfo {author} {\bibfnamefont {Jonathan~A.}\
  \bibnamefont {Jones}}, \bibinfo {author} {\bibfnamefont {Steven~D.}\
  \bibnamefont {Karlen}}, \bibinfo {author} {\bibfnamefont {Arzhang}\
  \bibnamefont {Ardavan}}, \ and\ \bibinfo {author} {\bibfnamefont {John
  J.~L.}\ \bibnamefont {Morton}},\ }\bibfield  {title} {\enquote {\bibinfo
  {title} {Magnetic field sensors using 13-spin cat states},}\ }\href@noop {}
  {\bibfield  {journal} {\bibinfo  {journal} {Phys. Rev. A}\ }\textbf {\bibinfo
  {volume} {82}},\ \bibinfo {pages} {022330} (\bibinfo {year}
  {2010})}\BibitemShut {NoStop}%
\bibitem [{\citenamefont {Schaffry}\ \emph {et~al.}(2010)\citenamefont
  {Schaffry}, \citenamefont {Gauger}, \citenamefont {Morton}, \citenamefont
  {Fitzsimons}, \citenamefont {Benjamin},\ and\ \citenamefont
  {Lovett}}]{SchaffryPRA}%
  \BibitemOpen
  \bibfield  {author} {\bibinfo {author} {\bibfnamefont {Marcus}\ \bibnamefont
  {Schaffry}}, \bibinfo {author} {\bibfnamefont {Erik~M.}\ \bibnamefont
  {Gauger}}, \bibinfo {author} {\bibfnamefont {John J.~L.}\ \bibnamefont
  {Morton}}, \bibinfo {author} {\bibfnamefont {Joseph}\ \bibnamefont
  {Fitzsimons}}, \bibinfo {author} {\bibfnamefont {Simon~C.}\ \bibnamefont
  {Benjamin}}, \ and\ \bibinfo {author} {\bibfnamefont {Brendon~W.}\
  \bibnamefont {Lovett}},\ }\bibfield  {title} {\enquote {\bibinfo {title}
  {Quantum metrology with molecular ensembles},}\ }\href@noop {} {\bibfield
  {journal} {\bibinfo  {journal} {Phys. Rev. A}\ }\textbf {\bibinfo {volume}
  {82}},\ \bibinfo {pages} {042114} (\bibinfo {year} {2010})}\BibitemShut
  {NoStop}%
\bibitem [{\citenamefont {Braunstein}\ \emph {et~al.}(1996)\citenamefont
  {Braunstein}, \citenamefont {Caves},\ and\ \citenamefont {Milburn}}]{bcm}%
  \BibitemOpen
  \bibfield  {author} {\bibinfo {author} {\bibfnamefont {S.~L.}\ \bibnamefont
  {Braunstein}}, \bibinfo {author} {\bibfnamefont {C.}~\bibnamefont {Caves}}, \
  and\ \bibinfo {author} {\bibfnamefont {G.~J.}\ \bibnamefont {Milburn}},\
  }\bibfield  {title} {\enquote {\bibinfo {title} {Generalized uncertainty
  relations: Theory, examples, and lorentz invariance},}\ }\href@noop {}
  {\bibfield  {journal} {\bibinfo  {journal} {Ann. Phys. - New York}\ }\textbf
  {\bibinfo {volume} {247}},\ \bibinfo {pages} {135} (\bibinfo {year}
  {1996})}\BibitemShut {NoStop}%
\bibitem [{\citenamefont {Luo}(2000)}]{LuoLMP}%
  \BibitemOpen
  \bibfield  {author} {\bibinfo {author} {\bibfnamefont {S.}~\bibnamefont
  {Luo}},\ }\bibfield  {title} {\enquote {\bibinfo {title} {Quantum fisher
  information},}\ }\href@noop {} {\bibfield  {journal} {\bibinfo  {journal}
  {Lett. Math. Phys.}\ }\textbf {\bibinfo {volume} {53}},\ \bibinfo {pages}
  {243} (\bibinfo {year} {2000})}\BibitemShut {NoStop}%
\bibitem [{\citenamefont {Petz}\ and\ \citenamefont {Ghinea}(2011)}]{petz}%
  \BibitemOpen
  \bibfield  {author} {\bibinfo {author} {\bibfnamefont {D.}~\bibnamefont
  {Petz}}\ and\ \bibinfo {author} {\bibfnamefont {C.}~\bibnamefont {Ghinea}},\
  }\bibfield  {title} {\enquote {\bibinfo {title} {Introduction to quantum
  fisher information},}\ }\href@noop {} {\bibfield  {journal} {\bibinfo
  {journal} {QP--PQ: Quantum Probab. White Noise Anal.,}\ }\textbf {\bibinfo
  {volume} {27}},\ \bibinfo {pages} {261--281} (\bibinfo {year}
  {2011})}\BibitemShut {NoStop}%
\bibitem [{\citenamefont {Barndorff-Nielsen}\ and\ \citenamefont
  {Gill}(2000)}]{BarndorffJPA}%
  \BibitemOpen
  \bibfield  {author} {\bibinfo {author} {\bibfnamefont {O.~E.}\ \bibnamefont
  {Barndorff-Nielsen}}\ and\ \bibinfo {author} {\bibfnamefont {R.~D.}\
  \bibnamefont {Gill}},\ }\bibfield  {title} {\enquote {\bibinfo {title}
  {Fisher information in quantum statistics},}\ }\href@noop {} {\bibfield
  {journal} {\bibinfo  {journal} {J. Phys. A: Math. Gen.}\ }\textbf {\bibinfo
  {volume} {33}},\ \bibinfo {pages} {4481--4490} (\bibinfo {year}
  {2000})}\BibitemShut {NoStop}%
\bibitem [{\citenamefont {Nagata}(2009)}]{Nagata09}%
  \BibitemOpen
  \bibfield  {author} {\bibinfo {author} {\bibfnamefont {K.}~\bibnamefont
  {Nagata}},\ }\bibfield  {title} {\enquote {\bibinfo {title} {Necessary and
  sufficient condition for {G}reenberger-{H}orne-{Z}eilinger diagonal states to
  be full $n$-partite entangled},}\ }\href@noop {} {\bibfield  {journal}
  {\bibinfo  {journal} {Int. J. Theor. Phys.}\ }\textbf {\bibinfo {volume}
  {48}},\ \bibinfo {pages} {3358--3364} (\bibinfo {year} {2009})}\BibitemShut
  {NoStop}%
\bibitem [{\citenamefont {Chen}\ \emph {et~al.}(2011)\citenamefont {Chen},
  \citenamefont {Chitambar}, \citenamefont {Modi},\ and\ \citenamefont
  {Vacanti}}]{arXiv:1005.4348}%
  \BibitemOpen
  \bibfield  {author} {\bibinfo {author} {\bibfnamefont {Lin}\ \bibnamefont
  {Chen}}, \bibinfo {author} {\bibfnamefont {Eric}\ \bibnamefont {Chitambar}},
  \bibinfo {author} {\bibfnamefont {Kavan}\ \bibnamefont {Modi}}, \ and\
  \bibinfo {author} {\bibfnamefont {Giovanni}\ \bibnamefont {Vacanti}},\
  }\bibfield  {title} {\enquote {\bibinfo {title} {Multipartite classical
  states and detecting quantum discord},}\ }\href@noop {} {\bibfield  {journal}
  {\bibinfo  {journal} {Phys. Rev. A}\ }\textbf {\bibinfo {volume} {83}},\
  \bibinfo {pages} {020101} (\bibinfo {year} {2011})}\BibitemShut {NoStop}%
\end{thebibliography}%
\end{document}